\documentclass[conference]{IEEEtran}
\IEEEoverridecommandlockouts
\usepackage{cite}
\usepackage{amsmath,amssymb,amsfonts}
\usepackage{algorithmic}
\usepackage{graphicx}
\usepackage{textcomp}
\usepackage{xcolor}
\def\BibTeX{{\rm B\kern-.05em{\sc i\kern-.025em b}\kern-.08em
    T\kern-.1667em\lower.7ex\hbox{E}\kern-.125emX}}
\usepackage{multirow}
\usepackage{cleveref}
\usepackage{pbalance}
\begin{document}

\title{Role-playing software architecture styles\thanks{Supported by the Centro de Investigación de Galicia ``CITIC'', funded by Xunta de Galicia and the European Regional Development Fund (grant ED431G 2019/01.}}

\author{Laura M.~Castro\\
\IEEEauthorblockA{\textit{Centro de Investigación en TIC (CITIC)} \\
\textit{Universidade da Coruña}\\
A Coruña, Spain \\
lcastro@udc.es -- https://orcid.org/0000-0002-3028-1523}
}

\maketitle

\begin{abstract}
Software Architecture, from definition to maintenance and evolution, is a complex aspect of software development and, consequently, a challenging subject when it comes to teaching it, and learning it.

Many research efforts have been devoted to designing teaching approaches, strategies and tools. Most of them, however, focus on the knowledge itself and the ways to convey it to students, rather than on the different learning styles of students themselves.

Teaching methods which predominantly rely on verbal and written communication, are very well aligned with some learning styles. However, students with learning styles that benefit more from physical activity or first-hand experience, need to defer to cognitive processes that are less natural to them. 

In this work, we propose an innovative use of role-playing as teaching strategy for architecture models of reference (i.e.~layered, pipe \& filter, client-server, etc.). This role-playing of different software architectures, in which students play the part of specific components in the system, intends to complement other classical teaching materials, such as in-person or recorded lectures, lab assignments, or development projects.

Addressing all learning styles within a classroom is key to ensure that we favour and foster the students' different learning processes, and give everyone an even playfield in which to best develop their capabilities as Software Architects.
\end{abstract}


\section{Introduction}

Teaching Software Architecture (SA) in a meaningful and effective way is difficult. Many reasons have been posed to explain this reality, although there seems to be some consensus around the idea that SA exposes students to concepts that have significantly greater scale than handled before, and which are not easy to reproduce within the classroom-and-academic-term context~\cite{VanDeursen2017591}. These concepts require the development of correspondingly increased abilities of abstraction, being also often the first time issues like performance or security are addressed~\cite{Galster2016356}. Last but not least, there are non-technical skills at play, such as communication~\cite{Lago200535} and decision-making~\cite{Capilla2020231}, that need to be fostered, too~\cite{Ferrari200932}.

In this paper, we propose a SA teaching approach driven by the co-existence of different learning profiles, as defined by David Kolb~\cite{kolb2007kolb,McLeod2017}, in the classroom. Taking into account that traditional activities such as lectures and development projects are more aligned with some of those learning profiles, we contribute a set of role-playing activities designed to improve student's understanding of different software architectures characteristics, strengths and weaknesses, that target those learning profiles which are usually forgotten.

\section{Role-playing as a teaching tool}

The use of participatory teaching methods features a wide range of options which are applicable in the context of computer science~\cite{Jones1987155}: brainstorming, directed dialogues, small discussion groups, debates, panel discussions, or role-playing, amongst others. From this range of choices, for this work we have decided to use role-playing as a teaching tool.

Role-playing, sometimes referred to as \emph{role-play simulation} in educational settings, is an experiential learning method in which learners are involved in a proposed scenario by representing an interacting part in it. The scenario is outlined by the teacher or professional, and while it must allow improvisation, it represents a safe and supportive environment where students will develop their own meaningful first-person experience.

Role-playing is widely acknowledged as a powerful technique across multiple avenues of training and education. There is previous experience in software engineering featuring role-playing, most often to simulate the conditions of an industrial environment~\cite{Ohlsson1995291}, of software product lines~\cite{Zuppiroli201213}, or for requirements engineering~\cite{Zowghi2003233}. In these cases, students are asked to play the part of stakeholders, from clients to the different technical roles within the development team.

However, the way we have chosen to use role-playing to teach SA is not by assigning human roles to students, but by assigning them the roles of software components, as we will discuss in Section~\ref{sec:roleplaying}. While ``traditional'' (i.e.~human-focused) role-playing has been used before in the context of SA teaching~\cite{Montenegro20171} (as a trade-off analysis method), we have only found one previous similar approach to ours (i.e.~software-focused): specifically, in the context of a programming course where students play the part of objects in order to grasp the concepts of object-orientation~\cite{Andrianoff2002121}.

\section{Role-playing architecture models}
\label{sec:roleplaying}

SA at University of A Coruña is a 6-ECTS (\emph{European Credit Transfer and accumulation System}~\cite{ects}) course for Software Engineering students. This translates in practice into 150 hours of work, typically spread amongst the 15 weeks of a term. Of the 10 weekly hours, 3 correspond to class and lab hours, and 7 to student autonomous work. In this course, students are exposed to architectural concepts and practices: from architectural patterns to non-functional requirements' analysis, the impact of the latter in the former, architectural representation and modelling, etc. The goal is to offer students a learning environment in which they can acquire the skills that allow them to carry out architectural tasks in the context of software development: component identification and characterisation, assignment of responsibilities, motivated election of communication and integration alternatives, and architectural evolution and maintenance.

\begin{figure}
  \centering
  \includegraphics[width=0.9\columnwidth]{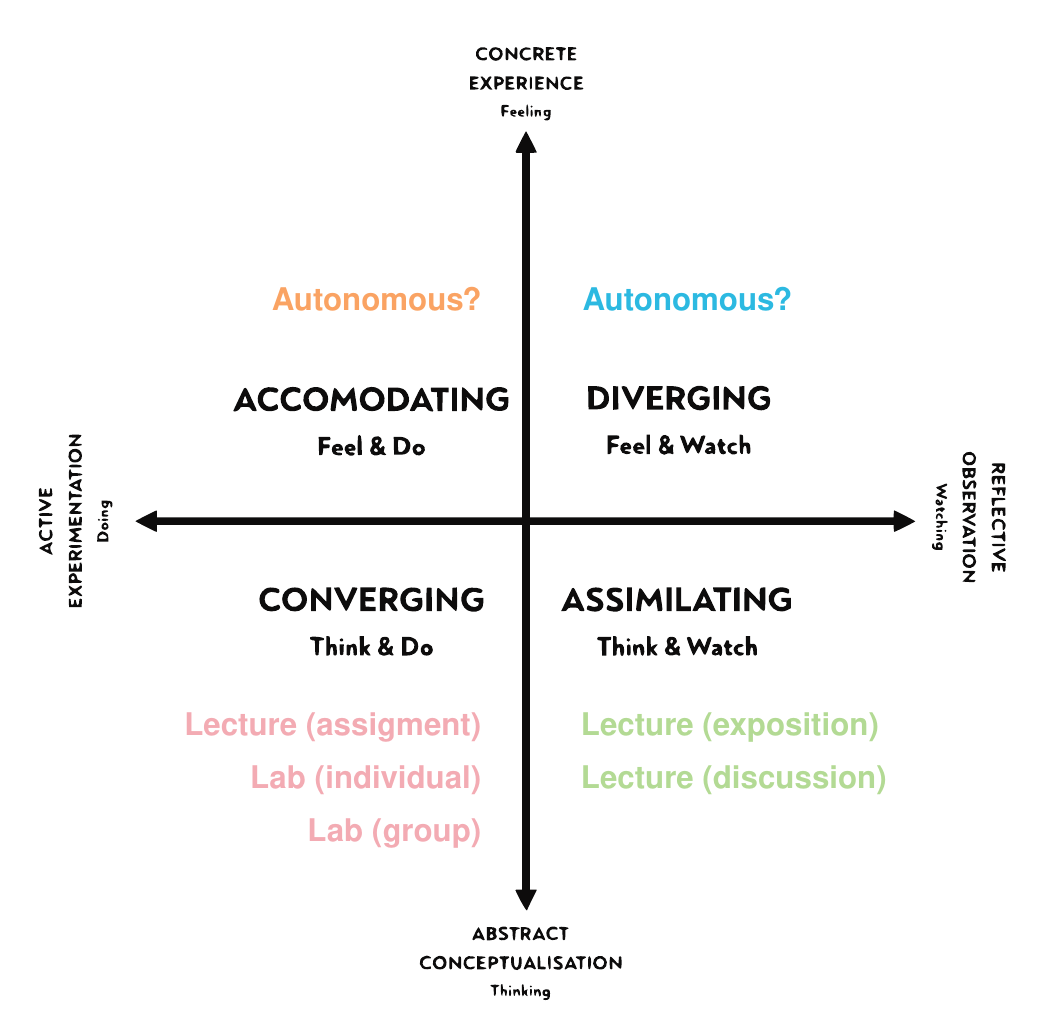}
  \caption{SA course at UDC: link between methodologies/learning styles}
  \label{fig:kolblearningstylesinAS}
\end{figure}

The teaching methodologies that were being used in the SA course at UDC were clearly aligned with some of Kolb's learning styles (cfg.~\ref{fig:kolblearningstylesinAS}). Lectures, and the materials provided (bibliographic references, articles, videos, assignments) work well for people which more prominent learning traits are watching and thinking, i.e.~those students that are reflective and take the most out of observation, and conceptualisation. Complementarily, practical work during lab sessions gave additional coverage to those most prominently guided by active experimentation. However, those with \emph{accommodating} and \emph{diverging} learning styles, in which experimentation or observation would most benefit from combination with concrete experience of the subject at hand, are entirely left up to self-manage their needs in their autonomous work.

Our proposal to address this situation was to role-play 
different architectures of reference,
meaning that \textbf{scenarios were designed to provide students with a concrete experience of the structure, advantages and disadvantages of different architectural models}.

\subsection{Role-playing a layered architecture}

The instructions given for the first role-playing scenario are shown in \Cref{tab:layers}. Students were made to form systems with different number of layers, but the same functionality. The expected outcomes are first-hand experiences of:

\begin{itemize}
  \item Layers reuse from one system to the next, increased when responsibilities are specific rather than broad.
  \item Limited impact of changes, affecting to single layers (or adjacent, at most).
  \item Request processing in two different moments at each layer (on the way \emph{in}, on the way \emph{out}, on both), and consequences (i.e.,~security).
  \item Structural limitations to performance, related to the number of layers.
\end{itemize}

\begin{table*}
  \caption[Roles to play in a layered architecture]{Roles to play in a layered architecture\textsuperscript{a}}
  \label{tab:layers}
  \footnotesize
  \centering
  \begin{tabular}{cp{0.6\textwidth}}
  \textsf{Role: layers} & \textsf{Instructions} \\\hline \hline
  \textit{You are the input layer} & Receive requests from the client (teacher) and pass them along after processing. Do not take another request until you have received the reply to the last. Give the reply back to the client. \\
  \textit{You are an intermediate layer} & Take requests only from your previous, pass them along after processing to your next. Do not take another request until you have received the reply to the last. Give the reply back only to your previous. \\
  \textit{You are the last layer} & Take requests only from your previous, and process it. Do not take requests until you are done. Give the reply back only to your previous. \\
  \end{tabular}
  \mbox{} \\[0.25cm]
  {\scriptsize\textsuperscript{a} Due to space limitations, these role character description cards are not as detailed as the ones students' received, but instead some abstracted, summarized versions.}
\end{table*}

\subsection{Role-playing a pipe \& filter architecture}

The instructions given for role-playing a pipeline are shown in \Cref{tab:pipelines} (\cpageref{tab:pipelines}). Groups of students were asked to define their own filters, and design their own pipeline. Expected outcomes were first-hand experience of:

\begin{itemize}
  \item Filters reuse from one pipeline to the next, increased when responsibilities are specific rather than broad.
  \item Limited impact of changes, affecting to single filters (or adjacent, at most).
  \item Advantage to performance given by structural parallelism, increased by the possibility of duplicating filter stages when one is a bottleneck.
  \item Differentiated input and output points in the system.
  \item One-way path for requests and consequences (i.e.~error handling).
\end{itemize}

\begin{table*}
  \caption{Roles to play in a pipeline architecture}
  \label{tab:pipelines}
  \footnotesize
  \centering
  \begin{tabular}{cp{0.6\textwidth}}
  \textsf{Role: filters} & \textsf{Instructions} \\\hline\hline
  \textit{You are the input filter} & Receive requests from the client (teacher) and pass them along after processing. Take a new request as soon as you have passed along the previous. \\
  \textit{You are an intermediate filter} & Take requests only from your previous, pass them along after processing to your next. Take new requests only after passing along the previous. \\
  \textit{You are the last filter} & Take requests only from your previous, process and reply to the expected output. \\
  \end{tabular}
\end{table*}

\subsection{Role-playing a client-server architecture}

\Cref{tab:client-server} (\cpageref{tab:client-server}) shows the instructions given for the client-server scenario, the first of the distributed architectural models. Students showcased the creation and evolution of several systems. The fist-hand experiences outcomes were:

\begin{itemize}
  \item Independence of services, in terms of development, availability, etc.
  \item Robustness: to the ability to have a working system as soon as the first service, and the directory, are operational, service independence adds the ability to keep in operation even if one or several services are down.
  \item Directory component as single point of failure, that can be mitigated by having different directories (opens the door for offering tailored service catalogues to distinct client profiles).
\end{itemize}

\begin{table*}
    \caption{Roles to play in a client-server architecture}
    \label{tab:client-server}
    \footnotesize
    \centering
    \begin{tabular}{cp{0.75\textwidth}}
    \textsf{Role: clients} & \textsf{Instructions} \\\hline\hline
    \textit{You are a client} & Your job is to submit requests for the services offered by the directory. You may submit one or many requests at a time, to the same or different services, whenever you want. \\ 
    \textsf{Role: directory} & \\\hline
    \textit{You are the directory} & Your job is to receive the requests from the clients and pass them along to one of the instances of the corresponding service. As soon as you have passed a request along, you can take the next one right away. \\ 
    \textsf{Role: services} & \\\hline
    \textit{You are a service} & Your job is to receive client requests forwarded by the directory and produce a reply for them. Do not take requests from clients directly. Do not take a new request until you are done with processing the previous. \\
    \end{tabular}
\end{table*}

\subsection{Role-playing a leader-follower architecture}

The instructions given for role-playing a leader-follower architecture are shown in \Cref{tab:leader-follower}. Students played the dynamic evolution during operation of several systems. Expected outcomes were first-hand experience of:

\begin{itemize}
  \item Independence of workers in terms of availability.
  \item Robustness: ability to stay operational even if some workers go down.
  \item Scalability: ability to elastically react to demand by starting more workers.
  \item Resource consumption optimization: ability to elastically react to demand by stopping idle workers.
  \item Leader as single point of failure.
\end{itemize}

\begin{table*}
    \caption{Roles to play in a leader-follower}
    \label{tab:leader-follower}
    \footnotesize
    \centering
    \begin{tabular}{cp{0.75\textwidth}}
    \textsf{Role: clients} & \textsf{Instructions} \\\hline\hline
    \textit{You are a client} & Your job is to submit requests to the system, via the leader. You may submit one or many requests at a time, whenever you want. \\
    \textsf{Role: leader} & \\\hline
    \textit{You are the leader} & Your job is to receive the requests from the clients and pass them along to one of the workers. You have to decide how to choose such worker (randomly, round-robin, etc). You may create a new worker if you find that all of them are busy. As soon as you have passed a request along, you can take the next one right away. You may stop a number of workers if you find that most of them are often idle. You may send the request to several workers. You may split the request into several workers. \\
    \textsf{Role: worker} & \\\hline
    \textit{You are a worker} & Your job is to process leader' requests. Do not take requests from anyone else. Do not take a new request until you are done with processing the previous. \\
    \end{tabular}
\end{table*}

\subsection{Role-playing a peer-to-peer architecture}

The instructions given for the last role-playing scenario are shown in \Cref{tab:p2p}. Part of the students were made to form one network of peers, while the rest were to perform as clients. Expected experiences-as-outcomes were:

\begin{itemize}
  \item Need to embed both management logic and business logic into each peer (as opposed to the leader-follower architecture, in which these responsibilities are assigned to distinct components).
  \item Need to manage life of requests, to ensure clients get some reply even when the network is under heavy load.
  \item Difficulties of implementing security measures to differentiate malicious from faulty or unavailable peers.
  \item Absence of coordination (impact on scalability, self-load balance).
  \item No single point of failure: maximum availability.
\end{itemize}

\begin{table*}
    \caption{Roles to play in a peer-to-peer architecture}
    \label{tab:p2p}
    \footnotesize
    \centering
    \begin{tabular}{cp{0.75\textwidth}}
    \textsf{Role: clients} & \textsf{Instructions} \\\hline\hline
    \textit{You are a client} & Your job is to submit requests to the system, using any of the peers. You may submit one or many requests at a time, whenever you want. \\
    \textsf{Role: peer} & \\\hline
    \textit{You are a peer} & Your job is to receive the requests (which may come from other peers or from clients) and either handle them and produce a reply, or pass them along to one or several of your neighbours. If you produce a reply, give it back to whoever sent it to you. As soon as you have processed/passed a request along, you can take the next one right away. Periodically, you need to revisit your neighbourhood, look for new peers, dismiss others. \\
    \end{tabular}
\end{table*}

\section{Discussion}

Literature shows that multi-role projects are an effective teaching strategy~\cite{Warin2016137}. It also reveals that role-playing affects three areas that support student learning and engagement, namely: identified personalized learning, deepened content understandings, and enhanced collaboration skills~\cite{Toth2015386}.

However, empirical evaluation also reports that while many students feel that the infusion of role-playing aspects into the courses supported their learning and engagement, some other students do not~\cite{Toth2015386}. This is coherent with the fact that certain teaching activities resonate better with certain learning profiles, and may also explain our quantitative evaluation. 

In any case, we consider this does not pose a threat to the validity of our efforts, since our role-playing activities aim precisely to provide an advantage to those learning profiles that benefit the less from the rest of teaching methodologies already present. There are a couple of factors that may influence how impactful this is when evaluating an initiative such as ours in the context of a whole class or course:

\begin{itemize}
    \item On the one hand, given that experimental learning experiences that address all learning styles are not commonplace, it makes sense to assume that students, especially by the time they reach the university level, have already adapted to the lack of methods and activities that best resonate with them, in the case of the less-often addressed learning profiles.
    \item On the other hand, given that we do not know how the distribution of different learning profiles amongst students in general, and neither amongst our CS students at UDC, it may also be the case their numbers are not significant when considered as part of the whole class. 
\end{itemize}

\section{Conclusions}

In this paper, we have faced the challenge of extending the teaching methodologies that were being applied in an undergraduate SA course. The goal was to take into account the different learning profiles that students have. In doing so, we aim to provide a more fair learning environment, where every individual has opportunities to connect with the learning experiences in the most meaningful and effective way to them. 

Given that the teaching methods that were in place were found to be miss-aligned when it came to students with \emph{accommodating} or \emph{diverging} learning styles (following Kolb's nomenclature), the way we have addressed this challenge has been by incorporating role-playing in an innovative way. We have designed a set of scenarios where students play the part of components in different architectural models. Such role-playing game provides concrete experiences of the structure, advantages, and disadvantages of different architectural models.


\end{document}